\documentclass[12pt]{article} 

\usepackage{color}
\definecolor{darkblue}{rgb}{0.1,0.1,.7}
\usepackage[dvips, colorlinks, linkcolor=darkblue, citecolor=darkblue, urlcolor=darkblue, linktocpage]{hyperref}
\usepackage[]{amsmath}
\usepackage[]{graphicx}
\usepackage[]{latexsym}
\usepackage{geometry}
\usepackage{amscd}
\usepackage[all,cmtip]{xy}
\usepackage[square, comma, sort&compress,numbers]{natbib}
\numberwithin{equation}{section}
\newcommand{\reef}[1]{(\ref{#1})}
\def\beq{\begin{equation}} 
\def\eeq{\end{equation}} 
\def\beqrr{\begin{array}} 
\def\eeqrr{\end{array}} 
\def\beqa{\begin{eqnarray*}} 
\def\eeqa{\end{eqnarray*}} 
 
\font\mybb=msbm10 at 12pt 
\def\bb#1{\hbox{\mybb#1}} 
 
\def\bR {\bb{R}}

\def\bM {\bb{M}}

\def\b1 {\bb{1}} 

\def\del {\partial} 
\def\nn{\nonumber}

\def\lesssim{\mathrel{\hbox{\rlap{\hbox{\lower4pt\hbox{$\sim$}}}\hbox{$<$}}}} 
\def\gtrsim{\mathrel{\hbox{\rlap{\hbox{\lower4pt\hbox{$\sim$}}}\hbox{$>$}}}} 
 
\def\calN {{\cal N}} 
\def\calO {{\cal O}}

\newcommand{\hhref}[1]{\href{http://arxiv.org/abs/#1}{arXiv:#1}}
\newcommand{\CG}{{$SO(d+1,1)$}}

\numberwithin{equation}{section}
\newcommand{\beg}{\begin{gather}} 
\newcommand{\eeg}{\end{gather}} 
\newcommand{\bem}{\begin{multline}} 
\newcommand{\eem}{\end{multline}} 

\begin{document} 

\vspace*{-.6in} \thispagestyle{empty}
\begin{flushright}
LPTENS--11/37
\end{flushright}
\vspace{.2in} {\Large
\begin{center}
{\bf Spinning Conformal Blocks}
\end{center}}
\vspace{.2in}
\begin{center}
{\large Miguel S. Costa$^{a}$, Jo\~ao Penedones$^{b}$,  David Poland$^{c,d}$, and Slava Rychkov$^{e}$}
\\
\vspace{.3in} 
\emph{${}^a\,$
Centro de F\'\i sica do Porto\\
Departamento de F\'\i sica e Astronomia\\
Faculdade de Ci\^encias da Universidade do Porto\\
Rua do Campo Alegre 687,
4169--007 Porto, Portugal}
\\
\vspace{.2in} 
\emph{${}^b\,$Perimeter Institute for Theoretical Physics\\ Waterloo, Ontario N2L 2Y5, Canada}
\\
\vspace{.2in} 
\emph{${}^c\,$Jefferson Physical Laboratory, Harvard University\\Cambridge, Massachusetts 02138, USA}
\\
\vspace{.2in}
\emph{${}^d\,$School of Natural Sciences, Institute for Advanced Study\\ Princeton, New Jersey 08540, USA}
\\
\vspace{.2in} 
\emph{${}^e\,$Laboratoire de Physique Th\'{e}orique, \'{E}cole Normale Sup\'{e}rieure,\\
and Facult\'{e} de Physique, Universit\'{e} Pierre et Marie Curie (Paris VI),
France}

\end{center}

\vspace{.2in}

\begin{abstract}
For conformal field theories in arbitrary dimensions, we introduce a method to derive the conformal blocks corresponding to the exchange of a traceless symmetric tensor appearing in four point functions of operators with spin. Using the embedding space formalism, we show that one can express all such conformal blocks in terms of simple differential operators acting on the basic scalar conformal blocks. This method gives all conformal blocks for conformal field theories in three dimensions.  We demonstrate how this formalism can be applied in a few simple examples. 
\end{abstract}
\vspace{.2in}
\begin{center}
{\it\large Dedicated to the memory of Francis Dolan}
\end{center}
\vspace{.3in}
\hspace{0.7cm} September 2011

\newpage

\setcounter{page}{1}

\tableofcontents

\vfill\eject

\section{Introduction}

Conformal field theories (CFTs) play an important role throughout theoretical physics.  Their use ranges from describing string worldsheets in two dimensions, to condensed matter systems in three dimensions, to having numerous phenomenological applications in beyond the standard model physics in four dimensions.  Moreover, in any number of dimensions they help to shed light on fundamental questions about quantum gravity and effective field theories, via the AdS/CFT correspondence.  However, while it is possible to find solvable CFTs in two dimensions (where one can use the full Virasoro algebra) or in highly supersymmetric theories, more generally very few tools exist to learn about the behavior of concrete CFTs.

Recently some progress has been made at pursuing the approach called the `conformal bootstrap'~\cite{FGG-bootstrap,pol}, which aims to learn about CFTs by understanding the constraints imposed by associativity of the operator product expansion (OPE).  In~\cite{us,david,us-global,Vichi,Poland:2011ey} it was for example demonstrated that crossing symmetry of four point functions of scalars combined with unitarity leads to completely general bounds on CFT operator dimensions and OPE coefficients.  On the other hand, in~\cite{joao} it was shown that explicit solutions to the constraints of crossing symmetry can be constructed in large $N$ theories, where the solutions are in a one-to-one correspondence with local interactions in AdS.  Additional progress has been made at understanding the constraints of unitarity in this context~\cite{Fitzpatrick:2010zm}, as well as in exploring the recently found analogy between CFT correlation functions in the Mellin representation and scattering amplitudes~\cite{MackMellin,joaoMellin,Fitzpatrick:2011ia,Paulos:2011ie}.

In the above concrete implementations of the bootstrap program, an extremely important role is played by the functions referred to as the \emph{conformal blocks}.  These functions depend on the conformal cross-ratios $u,v$, and represent the contribution of a given conformal primary operator and all of its descendants to a CFT four point function.  As such, conformal blocks are in principle completely determined by the dimensions and spins of the external primary operators, as well as those of the exchanged primary operator.

So far, conformal blocks have been worked out only for external operators being Lorentz scalars, while the exchanged operator can be a traceless symmetric tensor of arbitrary spin $l$.  In even space-time dimensions, closed form expressions for the conformal blocks in terms of hypergeometric functions are known~\cite{DO1,DO2}.  On the other hand, in odd dimensions, less explicit but perhaps still useful expressions exist, either in terms of simple integrals, or in terms of a double power series expansion in $u$ and $1-v$ for $l=0$, combined with recursion relations for higher spins~\cite{DO1,DO3}.

For external fields of nonzero spin, however, conformal blocks have so far remained elusive. Yet it would be very interesting to know them, at least for situations where the external spins are equal to 1 or 2.  The bootstrap program could e.g. then be attempted for four point functions containing the stress tensor and/or global symmetry currents.\footnote{See however~\cite{Fortin:2011nq} for recent progress at understanding the \emph{superconformal blocks} of currents in four dimensional theories with $\calN=1$ supersymmetry, where currents reside in scalar multiplets.}  These fields, unlike the external scalars used so far~\cite{us,joao,david,us-global,Vichi,Poland:2011ey}, are conserved, so they have protected (and thus known) dimensions.  They can then for example be used to study any CFT with a given global symmetry.  This is particularly important if one is interested in CFTs without low-dimension scalars, as would be expected in CFT duals to pure quantum gravity in AdS.

The purpose of the present paper is to make progress on the problem of finding conformal blocks for four point functions of operators with nonzero spin, building on the results and formalism developed in~\cite{Emb}.  Our main result is an expression which allows us to compute any such conformal block in terms of derivatives acting on the basic scalar conformal blocks, provided that the exchanged field is a \emph{traceless symmetric tensor}. This limitation is due to the fact that only traceless symmetric tensors can appear in the OPEs of scalars; we will also comment on possible approaches to obtaining the higher spin conformal blocks for other possible exchanged operators (such as antisymmetric tensors).  To arrive at our result, we will make extensive use of the `index-free' embedding space formalism presented in~\cite{Emb}.

This paper is organized as follows.  In section~\ref{sec:scalarCB} we review the derivation of the conformal blocks for four point functions of scalar operators.  In section~\ref{sec:spin} we begin by presenting a method to express three point functions containing operators of arbitrary spin in terms of differential operators acting on the basic scalar-scalar-spin $l$ three point functions.  We then argue that applying the same differential operators to scalar conformal blocks will generate higher spin conformal blocks, and we discuss some simple examples and special cases of the application of this technique. We conclude in section~\ref{sec:concl}.

\section{Scalar Conformal Blocks Revisited} \label{sec:scalarCB}

\subsection{Conformal Blocks and the Operator Product Expansion}   \label{sec:CB-OPE}
In this paper we will focus on conformal field theories in $d\ge 3$ Euclidean dimensions, so that the conformal group is \CG.
All of our equations can be Wick-rotated to the Minkowski signature, paying attention to the $i\epsilon$ prescription. We will assume 
that the reader is familiar with the basics of the theory, see e.g.~\cite{DF}, chapter 4. 

Let us begin by considering a correlation function between four scalar primaries $\phi_i$ of dimension $\Delta_i$, which is fixed by conformal invariance to be of the form
\begin{equation}
\langle\phi_{1}(x_{1})\phi_{2}(x_{2})\phi_{3}(x_{3})\phi_{4}(x_{4})\rangle
=\left(  \frac{x^2_{24}}{x^2_{14}}\right)  ^{\frac12\Delta_{12}%
}\left(  \frac{x^2_{14}}{x^2_{13}}\right)  ^{\frac12\Delta_{34}}
\frac{G(u,v)}{(x^2_{12})^{\frac12(\Delta_{1}+\Delta_{2})}(x_{34}^2)^{\frac12(\Delta_{3}+\Delta_{4})%
}}\,, \label{eq:4pt}%
\end{equation}
where $x_{ij}\equiv x_i-x_j$, $\Delta_{ij}\equiv \Delta_i-\Delta_j$, and $G(u,v)$ is an arbitrary function of the conformally invariant cross-ratios
\begin{equation}
u=\frac{x_{12}^{2}x_{34}^{2}}{x_{13}^{2}x_{24}^{2}},\quad v=\frac{x_{14}%
^{2}x_{23}^{2}}{x_{13}^{2}x_{24}^{2}}.
\end{equation}

In general, $G(u,v)$ can be expanded in a basis of functions called {\it conformal blocks}.  To obtain these, we should first consider the operator product expansion (OPE)
\beq
\phi_1(x_1)\phi_2(x_2)=\sum_{\calO} \lambda_{12 \calO}  C(x_{12},\del_{x_2})^{e_1\ldots e_l} \calO_{e_1\ldots e_l}(x_2)\,,
\label{eq:12OPE}
\eeq
where the sum runs over all primary operators $\calO$ which appear in the $\phi_1 \times \phi_2$ OPE. For the OPE of two scalars, these operators are traceless symmetric tensors of an arbitrary spin $l$. The function $C(x_{12},\del_{x_2})$ is then fixed by conformal invariance in terms of the operator dimensions, while $\lambda_{12 \calO}$ is an undetermined numerical coefficient. Inserting this OPE as well as the analogous OPE for $\phi_3 \times \phi_4$ into the four point function, we obtain the representation
\beq
\langle\phi_{1}(x_{1})\phi_{2}(x_{2})\phi_{3}(x_{3})\phi_{4}(x_{4}%
)\rangle =
\sum_\calO \lambda_{12 \calO}\lambda_{34 \calO} W_\calO(x_1,x_2,x_3,x_4)\,,
\label{eq:TO}
\eeq
where the contribution of the operator $\calO$, which we call a {\it conformal partial wave}, is given by
\begin{align}
W_\calO(x_1,x_2,x_3,x_4)&=
C(x_{12},\del_{x_2})^{e_1\ldots e_l} C(x_{34},\del_{x_4})^{f_1\ldots f_l} 
\langle \calO_{e_1\ldots e_l}(x_2) \calO_{f_1\ldots f_l}(x_4)\rangle\,
\nn \\
&= \left(  \frac{x^2_{24}}{x^2_{14}}\right)  ^{\frac12\Delta_{12}%
}\left(  \frac{x^2_{14}}{x^2_{13}}\right)  ^{\frac12\Delta_{34}}%
\frac{G_\calO(u,v)}{(x^2_{12})^{\frac12(\Delta_{1}+\Delta_{2})}(x_{34}^2)^{\frac12(\Delta_{3}+\Delta_{4})\,%
}}\,.
\label{eq:CBdef}
\end{align}
This equation defines a function $G_\calO(u,v)$ called a \emph{conformal block}.\footnote{Note that some authors also refer to $G_\calO$ as a conformal partial wave.  However, it seems more natural to us to apply the latter name to the whole function $W_\calO$.} We will comment in section \ref{sec:casimir} below why it is possible to express $W_\calO$ in the same form as Eq.~(\ref{eq:4pt}), via a function of cross ratios.

Comparing with Eq.~\reef{eq:4pt}, we obtain
\beq
G(u,v)= \sum_\calO \lambda_{12 \calO}\lambda_{34 \calO} {G_\calO(u,v)}\,.
\label{eq:CBexp}
\eeq
Thus, we see that the expansion in conformal blocks allow us to obtain a representation of the otherwise arbitrary function $G(u,v)$ in terms of CFT data, namely the spectrum of operator dimensions appearing in the OPE and the corresponding OPE coefficients. This explains their fundamental importance. 

\subsection{Explicit Conformal Blocks}   \label{sec:expl}

The most direct definition of a conformal block comes from Eq.~\reef{eq:CBdef}.  In fact, it was by using this definition that Dolan and Osborn first computed closed form expressions for the conformal blocks of an arbitrary spin-$l$ primary in $d=2,4$ \cite{DO1}. The $d=4$ result is given by
\begin{align}
G_\calO(u,v)  & =\frac{(-)^{l}}{2^{l}}
\frac{z\bar{z}}{z-\bar{z}}\left[  \,k_{\Delta+l}(z)k_{\Delta-l-2}(\bar
{z})-(z\leftrightarrow\bar{z})\right]\,,\\
k_{\beta}(x)  &  \equiv x^{\beta/2}{}_{2}F_{1}\left(  \frac{\beta-\Delta_{12}}2,\frac{\beta+\Delta_{34}}2,\beta;x\right)\,, 
\label{eq:DO}%
\end{align}
where the variables $z,\bar{z}$ are related to $u,v$ via%
\begin{equation}
u=z\bar{z},\quad v=(1-z)(1-\bar{z})\,.
 \label{eq:uvzzbar}%
\end{equation}
Unfortunately, similarly explicit expressions are not yet known in $d=3$, but there do exist simple integral representations and power series expansions (see~\cite{DO1} and more recently~\cite{DO3}) which may prove useful.

\subsection{Conformal Block Normalization}   \label{sec:boundary}

Next, we will explain the normalization of the
conformal blocks \reef{eq:DO}  by making a direct comparison with the OPE.  To begin, we will require that two point functions of primaries have a unit normalization.  For scalars, this is given by
\beq
\langle\phi_i(x) \phi_j(0)\rangle = \frac{\delta_{ij}}{(x^2)^{\Delta_i}}\,,
\eeq
while for tensors a convenient way to express the unit normalization condition is by contracting with two constant traceless symmetric rank $l$ tensors $K$ and $K'$:
\begin{align}
\langle K\cdot \calO(x)\, K'\cdot \calO(0)\rangle &= \frac{1}{(x^2)^\Delta} K^{a_1\ldots a_l} K'^{b_1\ldots b_l} I_{a_1b_1}(x)\ldots I_{a_lb_l}(x)\,, \\
I_{ab}(x)&\equiv\delta_{ab}-2\frac{x_a x_b}{x^2}\,.
\end{align}
The OPE coefficient $\lambda_{12\calO}$ is then defined via the three point function
\begin{align}
\langle\phi_{1}(x_{1})\phi_{2}(x_{2})K\cdot \calO(x_{3})\rangle 
&=
\frac{\lambda_{12{\calO}}\ K_{a_1\ldots a_l} Y^{a_{1}}\ldots Y^{a_l}}
{
(x^2_{12})^{\frac12(\Delta_{1}+\Delta_{2}-\Delta+l)}
(x^2_{23})^{\frac12(\Delta_{2}+\Delta-\Delta_{1}-l)}
(x^2_{13})^{\frac12(\Delta_{1}+\Delta-\Delta_{2}-l)}
}\,,
\label{eq:3ptl2}
\\[3pt]
Y^{a}  &\equiv \frac{x_{13}^{a}}{x_{13}^{2}}-\frac{x_{23}^{a}}{x_{23}^{2}%
}\,.
\end{align}
These definitions then correspond to the following normalization of the leading term in the OPE \reef{eq:12OPE}:
\beq
\phi_1(x_1)\phi_2(x_2)\underset{x_{12}\to 0}{\sim}
\frac  {\lambda_{12 \calO} }{(x^2_{12})^{\frac12(\Delta_{1}+\Delta_{2}-\Delta+l)}}\,
x_{12}^{a_1}\ldots  x_{12}^{a_l}\, \calO_{a_1\ldots a_l}(x_2)\,.
\label{eq:OPEnorm}
\eeq
An analogous expression holds when $x_{34}\to0$ in the $\phi_3 \times \phi_4$ OPE. Comparing with Eq.~\reef{eq:CBdef}, we see that the conformal blocks should have the asymptotic behavior
\beq
G_\calO(u,v)\underset{x_{12},x_{34}\to 0}{\sim}
\frac{ x_{12}^{a_1}\ldots  x_{12}^{a_l}\, \Pi_{a_1\ldots a_l}^{b_1\ldots b_l}\,
I_{b_1c_1}(x_{23})\ldots I_{b_lc_l}(x_{23})\, 
x_{34}^{c_1}\ldots  x_{34}^{c_l}} 
{
(x^2_{12} x^2_{34})^{\frac 12(-\Delta+l)} 
(x^2_{23})^{\Delta}
}
\,.
\label{eq:asGO1}
\eeq
Here $\Pi_{a_1\ldots a_l}^{b_1\ldots b_l}$ is the projector onto the space of traceless symmetric tensors of rank $l$. It is enough to apply it once, since contractions with $I_{bc}(x)$ tensors preserve the property of tracelessness.

Let us now use the formula\footnote{This formula follows from the theory of spherical harmonics and the fact that the Poisson kernel
is the generating function of Gegenbauer polynomials \cite{Erdelyi,SW}. The constant $c_{d,l}$ is easy to fix by considering the leading $x\cdot y$ asymptotics, for which the terms subtracting traces in the projector are irrelevant.} 
\begin{align}
x^{a_1}\ldots  x^{a_l}\, \Pi_{a_1\ldots a_l}^{b_1\ldots b_l}\, y_{b_1}\ldots  y_{b_l}&=c_{d,l}\,(x^2 y^2)^{l/2}\, C_l^{h-1}( \hat x\cdot \hat y)\,,\\
c_{d,l}&\equiv\frac{l!}{2^l(h-1)_l}\,,
\label{eq:Geg}
\end{align}
where  $h\equiv d/2$, $\hat x \equiv x/(x^2)^{1/2}$, $C_l^{h-1}(t)$ is a Gegenbauer polynomial, and $(n)_l=\Gamma(n+l)/\Gamma(n)$ is the Pochhammer symbol. The asymptotic behavior of Eq.~\reef{eq:asGO1} then becomes
\beq
G_\calO(u,v)\underset{x_{12},x_{34}\to 0}{\sim} 
c_{d,l}\,
\frac{(x^2_{12} x^2_{34})^{\frac 12\Delta} }
{(x^2_{23})^{\Delta}} \,
C_l^{h-1}
(\hat x_{12}\cdot I (x_{23}) \cdot \hat x_{34})\,.
\label{eq:asGO2}
\eeq
Now, in the limit $x_{12},x_{34}\to 0$ that we are considering the conformal cross ratios have the asymptotic behavior
\begin{align}
u=z\bar z\to 0,  \qquad & u \sim x^2_{12} x^2_{34}/x^4_{23}\,,\\
v\approx 1- z-\bar{z} \to 1, \qquad & v \sim 1+2u^{1/2} \hat x_{12}\cdot I (x_{23}) \cdot \hat x_{34}\,.
\end{align}
Substituting these into \reef{eq:asGO2}, we get the final result for the conformal block asymptotics:
\begin{align}
G_\calO(u,v)\underset{u\to0,v\to 1}{\sim} 
&c_{d,l}\,
u^{\frac 12\Delta} 
\,
C_l^{h-1}
\Bigl(\frac{v-1}{2u^{1/2}}\Bigr)\nn \\ 
\underset{z,\bar z\to0}{\sim}\ \  
&c_{d,l}\,
(z\bar z)^{\frac 12\Delta} 
\,
C_l^{h-1}
\Bigl(-\frac{z+\bar z}{2(z\bar z)^{1/2}}\Bigr)\,.
\label{eq:asGOfin}
\end{align}

This asymptotic behavior is valid in any dimension. The explicit $d=4$ conformal blocks \reef{eq:DO} behave in the same limit as 
\beq
G_\calO(u,v) \underset{z,\bar z\to0}{\sim} \frac{(-)^{l}}{2^{l}}
(z\bar z)^{\frac 12\Delta} \Bigl[\frac{\rho^{l/2+1}} {\rho-1} +( \rho\to \rho^{-1})\Bigr]\,,\qquad
\rho \equiv z/\bar z\,,
\eeq
which is identical to the $d=4$ case of \reef{eq:asGOfin} using the properties of the Gegenbauer polynomials.

Alternatively, and somewhat more easily, one can check the asymptotics by considering just one of the limits. For example, demanding that the partial wave $W_\calO$ is consistent with the $x_{34}\to0$ OPE, we see that the conformal block should behave as
\begin{align}
G_\calO(u,v) &\underset{x_{34}\to 0}{\sim}
\frac{ Y^{a_1}\ldots  Y^{a_l}\, \Pi_{a_1\ldots a_l}^{b_1\ldots b_l}\,
x_{34}^{b_1}\ldots  x_{34}^{b_l}} 
{
(x^2_{12} x^2_{34})^{\frac 12(-\Delta+l)} 
(x^2_{24} x_{14}^2)^{\frac12(\Delta-l)}
}
\nn \\
&\underset{x_{34}\to 0}{\sim} c_{d,l}\,
\left(\frac{x^2_{12} x^2_{34}}
{x^2_{24} x_{14}^2}\right)^{\Delta/2}  \,
C_l^{h-1}
(\hat Y\cdot \hat x_{34})\,.
\label{eq:as1}
\end{align}
Finally, by using
\beq
v \underset{x_{34}\to 0}{\sim} 1+2 Y\cdot x_{34}=
1+ 2 \frac{\hat Y\cdot \hat x_{34}}{u^{1/2}}\,,
\eeq
we obtain the same asymptotics as \reef{eq:asGOfin}.

\subsection{Casimir Differential Equation}   \label{sec:casimir}

An alternative and very efficient way to compute conformal blocks is based on using the Casimir differential equation. While we will not be relying on this method in this work (see footnote \ref{note:3}), we review it here for completeness.

Consider the conformal group generators $P_a$, $K_a$, $D$, $M_{ab}$, written collectively in the \CG\ notation as $M_{AB}$. Conformal invariance of the scalar four point function can be expressed by the equation
\beq
\left(\sum_{i=1}^4 M_{iAB}\right)\langle\phi_{1}(x_{1})\phi_{2}(x_{2})\phi_{3}(x_{3})\phi_{4}(x_{4})\rangle=0\,,
\label{eq:M4pt}
\eeq
where the generator $M_i$ acts on $\phi_i(x_i)$. The action of conformal generators on primary fields is given by well-known differential operators \cite{MS,FGG-book}, so that \reef{eq:M4pt} becomes a differential equation whose most general solution can be shown to have the form \reef{eq:4pt}.
 
 Conformal invariance of the OPE \reef{eq:12OPE} can be also expressed in this language. Acting on the LHS of this equation by $M_{1AB}+M_{2AB}$ corresponds to a certain differential operator with coefficients depending on $x_{1,2}$. Conformal invariance means that applying this differential operator to the RHS gives the same result as acting with $M_{AB}$ on the operator $\calO(x_2)$. 
 
 Using this observation, we can show that every $W_\calO$ in the decomposition \reef{eq:TO} can indeed be expressed in terms of a function of cross ratios as in \reef{eq:CBdef}. Applying $\sum_{i=1}^4 M_{iAB}$ to the first line of Eq.~\reef{eq:CBdef}, we can replace this differential operator by the sum of generators acting on $\calO(x_2)$ and $\calO(x_4)$, which gives zero since the two point function is conformal. Thus, every $W_\calO$ satisfies the same differential equation \reef{eq:M4pt} as the four point function itself, and should therefore have the same functional form.
 
Let us now apply the differential operator $(M_{1AB}+M_{2AB})(M_1^{AB}+M_2^{AB})$. By the same argument, the result will be equal to the one obtained by acting with $M_{AB}M^{AB}$ on the operator $\calO(x_2)$.
However, $M_{AB}M^{AB}$ is the quadratic Casimir of the conformal group, so that we have an eigenvalue equation
\beq
\frac 12 M_{AB}M^{AB} \calO(x_2)= C_{\Delta,l} \calO(x_2)\,,
\eeq
where the eigenvalue depends only on the dimension $\Delta$ and spin $l$ of the traceless symmetric primary:
\beq
C_{\Delta,l} =  \Delta ( \Delta - d) + l(l+d-2)  \,.
\label{eq:caseig}
\eeq
We conclude that the operator $\calO$ contribution in Eq.~\reef{eq:TO} satisfies the \emph{Casimir differential equation},
\beq
\Bigl[\frac 12 (M_{1AB}+M_{2AB})^2-C_{\Delta,l}\Bigr] W_\calO(x_1,x_2,x_3,x_4)=0\,.
\label{eq:casDE}
\eeq
This differential equation can then be rewritten\footnote{The actual computation is best performed by lifting to the embedding space where conformal generators act simply, see \cite{DO2}.} as an equation for $G_\calO(u,v)$, and solved in closed form when $d$ is an even integer \cite{DO2}. The results are consistent with those obtained by direct summation of the OPE series \reef{eq:CBdef}.

The Casimir differential equation is second order, so it is important to impose the correct boundary conditions in order to distinguish conformal blocks from other possible solutions. These boundary conditions are provided by the asymptotics \reef{eq:asGOfin}.

\section{Conformal Blocks of Operators with Spin} \label{sec:spin}
\subsection{Formulation of the Problem and the Basic Idea} \label{sec:basic}
 
In the previous section, we considered four point functions of scalar primary operators and expanded them into conformal blocks representing the exchange of an intermediate spin $l$ primary. 
We would now like to extend the conformal block construction to cases where the fields entering the correlator (the {\it external} fields) themselves have nonzero spin.
 Two particularly interesting examples are four point functions of the stress tensor and four point functions of conserved currents.
 
The definition of conformal blocks given in section \ref{sec:CB-OPE} generalizes easily to this situation.
In order to do this we have to generalize the OPE to cases where the operators $\phi_i$ have spin. This OPE will look like Eq.~\reef{eq:12OPE}, but the OPE coefficient function $C(x_{12},\del_2)$ will additionally carry external operator indices:
\beq
\phi_{1}^{\{a\}}(x_1)\phi_{2}^{\{b\}}(x_2)=\sum_\calO \lambda_{12 \calO}  C(x_{12},\del_{x_2})^{\{a,b,e\}}\, \calO_{\{e\}}(x_2)\,,
\label{eq:12ope}
\eeq
where we have used the shorthand notation $\{e\}\equiv e_1\ldots e_{l}$, etc.  We can then define partial waves $W_\calO$ analogously to the first line of Eq.~\reef{eq:CBdef}. 
Our goal is then to find explicit expressions for $W_\calO$, in the same way that Eqs.~\reef{eq:CBdef} and~\reef{eq:DO} solve this problem for scalar external fields.
 
One complication in the case of nonzero external spin is that there will be in general more than one partial wave corresponding to each exchanged field $\calO$. The number of partial waves will generically be given by the product $N_{12\calO} N_{34\calO}$, where $N_{12\calO}$ and $N_{34\calO}$ are the number of inequivalent ways in which $\calO$ can appear in the OPE $\phi_1\times \phi_2$ and $\phi_3\times \phi_4$, respectively. For external scalars we had $N_{12\calO}=N_{34\calO}=1$, but this is no longer true for  external operators with nonzero spin. 

The problem of finding conformal blocks for nonzero external spin then naturally splits into two steps.\footnote{\label{note:3}One could also in principle attempt a derivation based on the Casimir differential equation, which allows a straightforward generalization to the nonzero external spin case. However, proceeding this way one gets a coupled system of $N_{12\calO} \times N_{34\calO}$ second order equations which is nontrivial to solve. Even if the system could be solved, there would still remain a problem of establishing a correspondence between partial waves and OPEs. For these reasons we will not pursue this method here.} 
First, we need to classify all of the possible OPE structures.  Second, for each pairwise product of structures we need to perform the summation indicated in Eq.~\reef{eq:CBdef}. 

One can foresee that the second step is likely to be more complicated. After all, it was for many years that the OPE of two scalars and a spin $l$ field was known, but it was only in 2001 that Dolan and Osborn \cite{DO1} performed the summation and found the explicit conformal blocks. 

The main point of our paper is that the computation can be organized so that the second step is avoided altogether. The idea is that we can represent the OPE structures for external nonzero spin, Eq.~\reef{eq:12ope}, in terms of derivatives acting on the scalar OPE functions defined in Eq.~\reef{eq:12OPE}:
\beq
C(x_{12},\del_{x_2})^{\{a,b,e\}} =  D_{x_1,x_2}^{\{a,b\}}C(x_{12},\del_{x_2})^{\{e\}}\,.
\label{eq:Dab}
\eeq
Here $D$ is a differential operator, which creates open indices $\{a\}$ and $\{b\}$. Each different conformal structure in the LHS will have its own differential operator. Moreover, as indicated, this operator will be acting on the external field coordinates $x_1$ and $x_2$ only. This last property is crucial. It implies that once the OPEs $\phi_1\times\phi_2$ and $\phi_3\times\phi_4$ are plugged into the definition of partial waves, these extra differentiations 
factor out. As a result, partial waves of nonzero spin fields are reduced to derivatives of the known scalar partial waves:
\beq
 W^{\{a,b,c,d\}}_\calO(x_1,x_2,x_3,x_4)= D_{x_1,x_2}^{\{a,b\}} D_{x_3,x_4}^{\{c,d\}} 
 W_\calO(x_1,x_2,x_3,x_4)\,.
 \eeq
In the rest of the paper we will carry out this program in detail.

To summarize, the advantage of our approach is that it allows us to recycle a highly nontrivial computation---the summation of the double OPE series---which was already performed by Dolan and Osborn in the external scalar case. Unfortunately, there is a flipside to that coin: the only exchanged fields which can be treated by our method are those which are already present for the external scalars, which are traceless symmetric tensors of an arbitrary rank $l$. Partial waves corresponding to the exchange of antisymmetric tensors or fields of mixed symmetry cannot be found this way. We will comment on this limitation in more detail below.
  
\subsection{Classifying Three Point Functions} \label{sec:3pt}
 
The first step of the program that we outlined above consists in classifying conformally invariant OPEs \reef{eq:12ope},
and then representing them as derivatives of the scalar-scalar-spin $l$ OPE as indicated in Eq.~\reef{eq:Dab}.
As is well known, OPEs in conformal field theory are in one-to-one correspondence with conformally invariant three point functions.  Indeed, we used this fact when we discussed the normalization of the conformal blocks in section \ref{sec:boundary}.  Thus we have an equivalent problem: classify conformally invariant spin $l_1$-spin $l_2$-spin $l$ three point functions
and represent all of them via differential operators acting on the basic scalar-scalar-spin $l$ three point functions given in Eq.~\reef{eq:3ptl2},
\beq
\langle\phi_{1}^{\{a\}}(x_1)\,\phi_{2}^{\{b\}}(x_2)\, \calO^{\{e\}}(x_3)\rangle = D_{x_1,x_2}^{\{a,b\}} 
\langle\phi_{1}(x_1)\,\phi_{2}(x_2)\, \calO^{\{e\}}(x_3)\rangle\,.
\label{eq:3ptder}
\eeq

\subsubsection{Three Point Functions and the OPE}

The problem of classifying three point functions of arbitrary spin $l$ fields has previously been discussed several times 
in the literature \cite{Mack,Zaikov,OP}, as well as more recently in \cite{Emb}.
The results can be compactly presented by using a notation where fields are contracted with auxiliary $d$-dimensional polarization vectors \cite{Todorov},
\beq
 \phi(x;z)\equiv \phi_{a_1\ldots a_{l}}(x) z^{a_1}\ldots z^{a_l}\,.
 \eeq 
The most general three point function in physical space can then be written in the form \cite{Emb}
 \begin{multline}
\langle \phi_1(x_1;z_1)\phi_2(x_2;z_2) \phi_3(x_3 ;z_3)\rangle
= 
\frac{t\left(\tilde x_{12}, \tilde z_1,\tilde z_2 ,z_3\right)
}
{(x^2_{12})^{\frac12(\tau_1 + \tau_2 -\tau_3)}
   (x^2_{13})^{\frac12(\tau_1 + \tau_3 -\tau_2)} 
   (x^2_{23})^{\frac12(\tau_2 + \tau_3 -\tau_1)}}+O(z_i^2)\,,
\label{eq:3ptphys}
\end{multline}
where $\tau_i \equiv \Delta_i+l_i$, 
 \beq
 x_{ij}\equiv x_i-x_j,\quad \tilde{x}_{12} \equiv x_{13} \,x_{23}^2 - x_{23}\,x_{13}^2,\quad  \tilde z_1 \equiv I(x_{13})\,z_1,\quad \tilde z_2 \equiv I(x_{23})\, z_2\,, 
\eeq
and $t(x,z_1,z_2,z_3)$ is an arbitrary $O(d)$-rotational invariant homogeneous polynomial of degree $(l_1+l_2+l_3,l_1,l_2,l_3)$.

As indicated, Eq.~\reef{eq:3ptphys} only determines the LHS modulo $O(z_i^2)$ terms. We don't have to keep track of these terms because they are not independent -- they are fixed by the condition of tracelessness of the $\phi_i$'s. Explicitly, the full tensor can be recovered by using the formula
\cite{Todorov}
\beq
\phi_{a_1\dots a_l}(x) = \frac{1}{l! (h-1)_l}  D_{a_1} \cdots D_{a_l} [ \phi(x;z)+O(z^2)]\,,
\eeq
where the operator $D_a$ is given by
\beq
D_a=\left(h-1+z\cdot \frac{\partial}{\partial z} \right) \frac{\partial}{\partial z^a} -\frac{1}{2} z_a  
\frac{\partial^2\ }{\partial z \cdot \partial z}\,,
\label{eq:Tod}
\eeq
and we recall that $h\equiv d/2$.
The previous remark also means that polynomials $t(x,z_1,z_2,z_3)$ differing by an $O(z_i^2)$ amount really correspond to the same three point function. This needs to be kept in mind when counting the number of independent structures; see Eq.~\reef{eq:t} below.

The three point function \reef{eq:3ptphys} corresponds to an OPE whose leading singularity is expressed in terms of the same polynomial:
 \beq
\phi_1(x ,z_1)\phi_2(0,z_2) \underset{x\to 0}{\sim}(x^2) ^{-\frac 12(\tau_1+\tau_2-\tau_3)}  \phi_3(0,\del_{z_3})\, t(x,z_1,z_2,z_3)+O(z_1^2,z_2^2)\,.
\label{eq:3ptOPEphys}
\eeq
Notice that the form of this OPE is fixed by $O(d)$ invariance and the scaling dimensions of the fields. Eq.~\reef{eq:3ptphys} shows that any such $O(d)$ invariant OPE can be uniquely lifted to a fully conformally invariant three point function.

\subsubsection{Three Point Functions and the Embedding Formalism} \label{sec:3ptEmb}

The equations presented in the previous section solve the problem of classifying three point functions. Next we would like to understand how to represent them via differential operators acting on scalar-scalar-spin $l$ three point functions. 

A natural way to look for such a representation is to lift the whole problem to the $(d+2)$-dimensional \emph{embedding space}, where the conformal group is linearly realized. This approach to CFT computations goes back to the work of Dirac \cite{Dirac}, and has been used on and off since the 1970's \cite{Boulware,FGG-bootstrap,FGG-book} (see also \cite{CCP,Weinberg} for recent work). An extensive discussion can be found in \cite{Emb}, where we have developed an improved index-free version of the formalism, which uses polynomials in auxiliary vector variables to encode embedding space tensors. We will now give a brief review of the formalism and explain how the three point functions are lifted to the embedding space. The problem of finding differential operators will be treated in the next subsection.

In the embedding formalism, points $x\in \bR^d$ in physical space are put in a one-to-one correspondence with light rays through the origin of  the embedding space $\bM^{d+2}$. Lorentz transformations \CG\  of the null vectors $P\in \bM^{d+2}$ (where $P^2=0$) then generate conformal transformations. One goes back to the physical space by projecting onto the Poincar\'e section 
of the light cone
\beq
P_x= (P^+,P^-,P^a) =( 1, x^2, x^a)\,.
\label{eq:Px}
\eeq 
Primary fields of dimension $\Delta$ and spin $l$ are encoded into a field $\Phi(P,Z)$, polynomial in the $(d+2)$-dimensional polarization vector $Z$, such that
\beq
\Phi(\lambda P; \alpha Z + \beta P ) = \lambda^{-\Delta} \alpha^l \Phi(P;Z)\,.
\label{eq:EmbField}
\eeq
Here the scaling with $\lambda$ encodes the dimension of the field, while the scaling with $\alpha$ means that the polynomial has rank $l$ (since it was obtained by contracting a rank $l$ tensor $\Phi_{A_1\ldots A_l}$ with the polarization vector $Z^A$). The invariance under shifts by $\beta P$ is a consequence of the fact that the embedding space tensor $\Phi_{A_1\ldots A_l}$ is required to be transverse. It is important that this invariance can be required to hold identically in the correlation functions (as opposed to the possible weaker requirement of being satisfied modulo terms that vanish on the $P^2=0$ cone).

Using this formalism, the most general form of the embedding space three point function containing operators with spins $l_i$ and dimensions $\Delta_i$ can be written as
\beq
\left\langle \Phi_1(P_1;Z_1) \Phi_2(P_2;Z_2)  \Phi_3(P_3;Z_3)\right\rangle  =\sum_{n_{12},n_{13},n_{23}\ge 0} \lambda_{n_{12},n_{13},n_{23}}
\left[ {\scriptsize
\begin{array}{ccc}
\Delta_1 & \Delta_2 & \Delta_3 \\
l_1 & l_2 & l_3 \\
n_{23} & n_{13} & n_{12} \end{array} }
\right]+O(Z_i^2,Z_i\cdot P_i)\,.
\label{eq:embsum}
\eeq
Here the sum runs over all the possible elementary three point function structures (that we will give below), each with arbitrary numerical coefficients.  To shorten the notation we also suppress their dependence on $P_i$ and $Z_i$.
The ``$+O(Z_i^2,Z_i\cdot P_i)$" means that such terms are not independent and we do not have to keep track of them. This feature, discussed in detail in \cite{Emb}, is the hallmark and the great simplification that occurs in our version of the embedding formalism. It is the embedding space counterpart of not having to keep track of the $O(z_i^2)$ terms in Eq.~\reef{eq:3ptphys}.

The elementary three point function structures appearing in Eq.~\reef{eq:embsum} are characterized by a choice of three nonnegative integers $n_{ij}$. Additionally, these numbers are required to satisfy the constraints
\beq
m_1\equiv l_1-n_{12}-n_{13}\ge 0\, ,\ \ \ \ \ \ \ \ 
m_2\equiv l_2-n_{12}-n_{23}\ge 0\, ,\ \ \ \ \ \ \ \ 
m_3\equiv l_3-n_{13}-n_{23}\ge 0\,.
\label{eq:nij}
\eeq
The three point function structures themselves are given by \cite{Emb}
\begin{align}
\left[ {\scriptsize
\begin{array}{ccc}
\Delta_1 & \Delta_2 & \Delta_3 \\
l_1 & l_2 & l_3 \\
n_{23} & n_{13} & n_{12} \end{array} }
\right]\,
&\equiv
  \frac{V_1^{m_1} V_2^{m_2}  V_3^{m_3} H_{12}^{n_{12}} H_{13}^{n_{13}} 
  H_{23}^{ n_{23}} }{(P_{12})^{\frac12(\tau_1 + \tau_2 -\tau_3)}
   (P_{13})^{\frac12(\tau_1 + \tau_3 -\tau_2)} 
   (P_{23})^{\frac12(\tau_2 + \tau_3 -\tau_1)}}\,,\label{eq:general3pt}\\[3pt]
   \nn
   P_{ij} &\equiv -2 P_i\cdot P_j\,.
\end{align}
The basic building blocks $V_i$ and $H_{ij}$ entering this equation are defined as
\begin{align}
H_{ij} &\equiv - 2\big[ (Z_i \cdot Z_j )( P_i\cdot P_j) - (Z_i \cdot P_j )( Z_j\cdot P_i)\big]\,,
\\
V_{i,jk} &\equiv \frac{(Z_i \cdot P_j )(P_i\cdot P_k )- (Z_i \cdot P_k )( P_i\cdot P_j)}{(P_j \cdot P_k)}\,,
\label{eq:HV} 
\\
V_1 &\equiv V_{1,23} \,, \quad V_2 \equiv V_{2,31} \,, \quad V_3 \equiv V_{3,12}\,.
\end{align}
These particular elementary polynomials are chosen because they are explicitly transverse (they don't change under $Z_i\to Z_i+\beta P_i$); they are also normalized to have a simple scaling in $P_i$ and a simple projection to the physical space (see below).

The number of elementary three point functions is equal to the number of solutions to \reef{eq:nij}, and can be found in closed form:
\beq
N(l_1,l_2,l_3)=\frac{(l_1+1)(l_1+2)(3l_2-l_1+3)}{6} - \frac{p(p+2)(2p+5)}{24} - \frac{1-(-1)^p}{16} \,,
\label{N3pt}
\eeq
where we have ordered the spins $l_1\le l_2\le l_3$ and defined $p\equiv {\rm max} (0,l_1+l_2-l_3)$.

Finally, to have a complete description, we would like to know which physical space three point function (or, equivalently, OPE) corresponds to the embedding space correlator \reef{eq:general3pt}. As shown in \cite{Emb}, to go to physical space we have to replace
\beq
P_i\to P_{x_i},\qquad Z_i \to Z_{z_i,x_i}\equiv (0,2 x_i \cdot z_i, z_i)\,.
\label{eq:projEmb}
\eeq
Doing this substitution in \reef{eq:general3pt} leads to a physical space three point function of the form \reef{eq:3ptphys}
with the polynomial $t(x,z_1,z_2,z_3)$ given by
\beq
  (x^2 z_1\cdot z_3)^{n_{13}}(x^2 z_2\cdot z_3)^{n_{23}}
 (x^2 z_1\cdot z_2 -2x\cdot z_1\, x\cdot z_2 )^{n_{12}}\,
 (-x\cdot z_1)^{m_1} (-x\cdot z_2)^{m_2} (x\cdot z_3)^{m_3}\,.
\label{eq:t}
\eeq
This is clearly the most general $O(d)$ invariant polynomial of the required degree, modulo terms of $O(z_i^2)$.

The discussion in this section needs to be somewhat modified in three special situations: for parity odd three point functions, in $d = 3$ dimensions, and for conserved tensors. These exceptional cases will be discussed later, so as to not interrupt the main line of reasoning.

\subsection{Differential Representation of Three Point Functions}

The differential operator in Eq.~\reef{eq:3ptder} will transform under the conformal group with transformation properties dictated by the fact that it maps one conformally invariant object into another. Therefore, the problem of finding the differential representation~\reef{eq:3ptder} will be greatly simplified by lifting it to the embedding space, where the conformal group acts linearly on the coordinates. 

 \subsubsection{Elementary Differential Operators}

Using the embedding space three point functions given in the previous section, the embedding space version of \reef{eq:3ptder} can be written as
\begin{multline}
\left[ {\scriptsize
\begin{array}{ccc}
\Delta_1 & \Delta_2 & \Delta_3 \\
l_1 & l_2 & l_3 \\
n_{23} & n_{13} & n_{12} \end{array} }
\right]=\mathcal{D}\left(P_{i},Z_{i},\frac\del{\del P_{i}}, \frac\del{\del Z_{i}}\right)
\left[ {\scriptsize
\begin{array}{ccc}
\Delta_1' & \Delta_2' & \Delta_3 \\
0 & 0 & l_3 \\
0 & 0 & 0 \end{array} }
\right] +O(Z_i^2,Z_i\cdot P_i,P_i^2) \qquad (i=1,2)\,,
\label{eq:embD}
\end{multline}
where it is important that the differential operator $\mathcal{D}$ only contains the indicated coordinates and derivatives. The dimensions $\Delta_{1,2}'$ appearing on the right hand side can (and in general will) be different from $\Delta_{1,2}$. On the other hand, the quantum numbers of the third field remain the same. 

The operator $\mathcal{D}$ must satisfy a number of consistency conditions. The most important condition is that it should take terms which are $O(Z_i^2,Z_i\cdot P_i, P_i^2)$ to terms of the same kind. This condition can be stated equivalently by saying that it should act tangentially to the submanifold defined by the equations
\beq
Z_i^2=Z_i\cdot P_i=P_i^2=0\qquad(i=1,2)\,.
\label{eq:surf}
\eeq
The reason for this condition is that the correspondence between the embedding space and physical space fields, as expressed by Eq.~\reef{eq:projEmb}, corresponds to restricting to this submanifold. Thus, if we want Eq.~\reef{eq:embD} to correspond to a differential relation between physical space quantities after imposing this restriction, the differential operator must act within the submanifold.

The second condition is that $\mathcal{D}$ must map explicitly transverse functions to themselves. This is because all elementary three point functions are explicitly transverse.

The operator $\mathcal{D}$ must raise the degree in the $Z_1,Z_2$ variables from $0$ to $l_1,l_2$. We will therefore construct it as a composition of elementary operators which raise the degree one unit at a time. A candidate set of four simple first-order operators are
\beq
Z_i \cdot \frac{\partial}{\partial P_j}\qquad(i,j=1,2)\,.
\eeq
However, we must add extra terms to these operators in order to satisfy the just mentioned consistency conditions. 

The choice of these compensating terms turns out to be unique. The four first-order operators which satisfy all of the conditions can be written as
\begin{align}\label{eq:diffops}
D_{11} &\equiv (P_1 \cdot P_2) (Z_1 \cdot \frac{\partial}{\partial P_2}) - (Z_1 \cdot P_2) (P_1 \cdot \frac{\partial}{\partial P_2}) - (Z_1 \cdot Z_2) (P_1 \cdot \frac{\partial}{\partial Z_2}) + (P_1 \cdot Z_2) (Z_1 \cdot \frac{\partial}{\partial Z_2})\,, \nonumber
\\
D_{12} &\equiv (P_1 \cdot P_2) (Z_1 \cdot \frac{\partial}{\partial P_1}) - (Z_1 \cdot P_2) (P_1 \cdot \frac{\partial}{\partial P_1}) + (Z_1 \cdot P_2) (Z_1 \cdot \frac{\partial}{\partial Z_1})\,,
\end{align}
as well as two more with the roles of $1$ and $2$ interchanged,
\begin{align}\label{eq:diffops2}
D_{22} &\equiv (P_2 \cdot P_1) (Z_2 \cdot \frac{\partial}{\partial P_1}) - (Z_2 \cdot P_1) (P_2 \cdot \frac{\partial}{\partial P_1}) - (Z_2 \cdot Z_1) (P_2 \cdot \frac{\partial}{\partial Z_1}) + (P_2 \cdot Z_1) (Z_2 \cdot \frac{\partial}{\partial Z_1})\,, \nonumber\\
D_{21} &\equiv (P_2 \cdot P_1) (Z_2 \cdot \frac{\partial}{\partial P_2}) - (Z_2 \cdot P_1) (P_2 \cdot \frac{\partial}{\partial P_2}) + (Z_2 \cdot P_1) (Z_2 \cdot \frac{\partial}{\partial Z_2})\, .
\end{align}
In the chosen notation, acting with $D_{ij}$ on a correlator increases the spin at point $i$ by one unit and decreases the dimension at point $j$ by one unit. 

The fifth operator (zeroth-order) is multiplication by $H_{12}$, which trivially satisfies all of the conditions. This increases the spin and decreases the dimension by one unit at both points.
 
 \subsubsection{Recursion Relations and the Differential Basis}
\label{sec:recursion}

The five introduced operators satisfy a set of recursion relations which allow us to find a differential representation for an arbitrary elementary three point function. 

We first present two recursion relations which connect three point functions with $n_{23}=n_{13}=0$. The first relation has the form\footnote{Here and below we are dropping terms of $O(Z_i^2, Z_i \cdot P_i, P_i^2)$ (for $i=1,2$), which are sometimes generated by the action of $D_{ij}$.}
\begin{align}\label{eq:addV1}
&( \Delta_3+l_1+l_2 -2n_{12}-2)
 \left[ {\scriptsize
\begin{array}{ccc}
\Delta_1 & \Delta_2 & \Delta_3 \\
l_1 & l_2 & l_3 \\
0  & 0  & n_{12} \end{array} }
\right]  \\&= D_{12} \,
  \left[ {\scriptsize
\begin{array}{ccc}
\Delta_1  & \Delta_2+1 & \Delta_3 \\
l_1-1 & l_2 & l_3 \\
0  & 0  & n_{12} \end{array} }
\right] +
D_{11} \,
  \left[ {\scriptsize
\begin{array}{ccc}
\Delta_1+1  & \Delta_2  & \Delta_3 \\
l_1-1 & l_2 & l_3 \\
0  & 0  & n_{12} \end{array} }
\right] 
-(l_2 -n_{12})   \left[ {\scriptsize
\begin{array}{ccc}
\Delta_1    & \Delta_2   & \Delta_3 \\
l_1  & l_2  & l_3 \\
0  & 0  & n_{12}  +1\end{array} }
\right]\nonumber\,.
\end{align} 
The second one is obtained by interchanging $1\leftrightarrow 2$. These two relations can be used together to recursively generate three point functions with $n_{13}=n_{23}=0$ and arbitrary values of $l_1$ and $l_2$, starting from the $l_1=l_2=0$ seeds.  
Indeed, the first two terms on the right hand side involve lower spin three point functions. The third term has the same spin; however, it has a larger value of $n_{12}$, and its coefficient vanishes when $n_{12}$ takes the maximal value allowed by \reef{eq:nij}. Thus, three point functions can be computed starting from $n_{12}=\min(l_1,l_2)$ and going down.
This starting point is easily constructed using the trivial recursion relation
\begin{align}
 \left[ {\scriptsize
\begin{array}{ccc}
\Delta_1 & \Delta_2 & \Delta_3 \\
l_1 & l_2 & l_3 \\
n_{23} & n_{13} & n_{12} \end{array} }
\right]  = H_{12} \,
  \left[ {\scriptsize
\begin{array}{ccc}
\Delta_1+1 & \Delta_2+1 & \Delta_3 \\
l_1-1 & l_2-1 & l_3 \\
n_{23} & n_{13} & n_{12}-1 \end{array} }
\right] \, ,\label{eq:addH12}
\end{align}
 which does not change $n_{13}$ or $n_{23}$, but raises both spins by one unit.
Finally, to get the remaining three point functions we should use the recursion relation
\begin{align} 
\label{eq:addH13}
&m_3
 \left[ {\scriptsize
\begin{array}{ccc}
\Delta_1 & \Delta_2 & \Delta_3 \\
l_1  & l_2 & l_3 \\
n_{23} & n_{13}+1 & n_{12}  \end{array} }
\right] =-2 D_{12} \,
  \left[ {\scriptsize
\begin{array}{ccc}
\Delta_1  & \Delta_2+1 & \Delta_3 \\
l_1-1 & l_2 & l_3 \\
n_{23} & n_{13} & n_{12}  \end{array} }
\right]  \\&+ 
m_2 \,
  \left[ {\scriptsize
\begin{array}{ccc}
\Delta_1  & \Delta_2  & \Delta_3 \\
l_1  & l_2 & l_3 \\
n_{23} & n_{13}  & n_{12} + 1 \end{array} }
\right] 
+(\tau_1+\tau_3-\tau_2+2m_2-2m_3 -2 )  \left[ {\scriptsize
\begin{array}{ccc}
\Delta_1    & \Delta_2   & \Delta_3 \\
l_1   & l_2 & l_3 \\
n_{23} & n_{13} & n_{12}  \end{array} }
\right]\nonumber\,,
\end{align}
as well as the similar equation obtained by permuting $1\leftrightarrow2$. Here the values of the $m_i$ are given by (\ref{eq:nij}).
These relations allow us to recursively generate nonzero values of $n_{13}$ and $n_{23}$.

To summarize, these recursion relations show that it is possible to represent any three point function in the form \reef{eq:embD}.
To be more precise, we will get a sum of several terms on the RHS corresponding to several $\Delta_{1,2}'$ satisfying the constraint
\beq
\Delta_1'+\Delta_2'=\tau_1+\tau_2\,.
\eeq
This is a minor technical detail which was not mentioned before so as to avoid cluttering notation. It implies that, in general, a given spinning conformal block will be related to a sum of several scalar conformal blocks.

In fact, in practice the computation can be organized without using recursion relations.  Instead, we just have to find the change of basis between the standard three point function basis \reef{eq:general3pt} and the differential operator basis defined by
\begin{align}
 \left\{ {\scriptsize
\begin{array}{ccc}
\Delta_1   & \Delta_2 & \Delta_3 \\
l_1 & l_2 & l_3 \\
n_{23} & n_{13} & n_{12} \end{array} }
\right\} \equiv
H_{12}^{n_{12}}
{D}_{12}^{n_{13}}
{D}_{21}^{n_{23}}
{D}_{11}^{m_1}
{D}_{22}^{m_2}
  \left[ {\scriptsize
\begin{array}{ccc}
\Delta_1+m_1+n_{23} +n_{12} & \Delta_2+m_2+n_{13} +n_{12}
& \Delta_3 \\
0 & 0 & l_3 \\
0 & 0 & 0 \end{array} }
\right]\,,  \label{Dbasis}
\end{align}
where the $m$'s are again fixed by (\ref{eq:nij}).
Note that in this expression we have chosen a particular ordering for the operators. Different orderings can be related using the 
commutation relations
\begin{align}
\left[ D_{11},{D}_{22}\right]&= \frac{1}{2}H_{12}
\left(Z_1 \cdot \frac{\partial}{\partial Z_1}
-Z_2 \cdot \frac{\partial}{\partial Z_2}
+P_1 \cdot \frac{\partial}{\partial P_1}
-P_2 \cdot \frac{\partial}{\partial P_2}
\right)\,,\\
\left[ {D}_{12},{D}_{21}\right]&= \frac{1}{2}H_{12}
\left(Z_1 \cdot \frac{\partial}{\partial Z_1}
-Z_2 \cdot \frac{\partial}{\partial Z_2}
-P_1 \cdot \frac{\partial}{\partial P_1}
+P_2 \cdot \frac{\partial}{\partial P_2}
\right)\,.
\end{align}
All other commutators vanish, including 
$\left[ {D}_{ij},H_{12}\right]=0$ for all $i$ and $j$.

Acting with the differential operators in \reef{Dbasis}, the differential basis elements can be expanded in terms of the standard basis \reef{eq:general3pt}.
By inverting the matrix, we can then get a differential operator representation for all elements of the standard basis \reef{eq:general3pt}.

\subsubsection{Example}
\label{sec:example3pt}
We will now demonstrate how the developed formalism works in the case that $l_1=l_2=1$ and $\Delta_1 = \Delta_2$.  Assuming that $l=l_3\ge 2$, we have five possible structures in the standard basis: 
\begin{align}
  \left[ {\scriptsize
\begin{array}{ccc}
\Delta_1 & \Delta_1 & \Delta \\
1 & 1 & l \\
0 & 0 & 0 \end{array} }
\right]
 ,
\left[ {\scriptsize
\begin{array}{ccc}
\Delta_1 & \Delta_1 & \Delta \\
1 & 1 & l \\
1 & 0 & 0 \end{array} }
\right]  ,
\left[ {\scriptsize
\begin{array}{ccc}
\Delta_1 & \Delta_1 & \Delta \\
1 & 1 & l \\
0 & 1 & 0 \end{array} }
\right]  ,
\left[ {\scriptsize
\begin{array}{ccc}
\Delta_1 & \Delta_1 & \Delta \\
1 & 1 & l \\
1 & 1 &0 \end{array} }
\right] 
,
\left[ {\scriptsize
\begin{array}{ccc}
\Delta_1 & \Delta_1 & \Delta \\
1 & 1 & l \\
0 & 0 & 1 \end{array} }
\right]  \,, \label{5structures}
\end{align}
as well as the five analogous structures in the differential basis (\ref{Dbasis}). We will denote these structures respectively as $[I]$ and $\{I\}$, where $I=1,\ldots,5$ in the obvious notation.

Implementing the action of the differential operators on a computer, it is simple to compute the matrix expressing the differential basis in terms of the standard one: 
\beq
\{I\}=\sum_{J=1}^5 \,a_{IJ} \,[J]\,,
\label{eq:a}
\eeq
which can then be inverted to express the $[J]$'s in terms of the $\{I\}$'s.  The matrix inverse $(a_{IJ})^{-1}$ is then given by
\begin{align}
\frac{1}{\Delta(1-\Delta)} 
\left(
\begin{array}{ccccc}
 1 & 1 & 1 & 1 & \Delta -1 \\
 \frac{\Delta -l}{l} & \frac{-\Delta
   -l}{l} & \frac{\Delta -l}{l} &
   \frac{-\Delta -l}{l} & 1-\Delta 
   \\
 \frac{\Delta -l}{l} & \frac{\Delta
   -l}{l} & \frac{-\Delta -l}{l} &
   \frac{-\Delta -l}{l} & 1-\Delta 
   \\
 \frac{(l-\Delta )^2}{(l-1) l} &
   \frac{(l-\Delta ) (\Delta
   +l)}{(l-1) l} & \frac{(l-\Delta )
   (\Delta +l)}{(l-1) l} &
   \frac{(\Delta+l) ^2-4 \Delta}{(l-1) l} &
   \frac{(\Delta -1) (l-\Delta
   )}{l-1} \\
 0 & 0 & 0 & 0 & (1-\Delta ) \Delta
\end{array}
\right)\,.
\end{align}

\subsection{From Three Point Functions to Conformal Blocks}
\label{sec:main}

We will now explain the last step of the algorithm: how to use the differential representation of the three point functions in order to compute conformal partial waves for operators with spin. This part of the computation is also natural to perform in the embedding space. The lift of the scalar partial waves \reef{eq:CBdef} is given by
\beq
W_\calO(P_1,P_2,P_3,P_4)
= \left(  \frac{P_{24}}{P_{14}}\right)  ^{\frac12\Delta_{12}%
}\left(  \frac{P_{14}}{P_{13}}\right)  ^{\frac12\Delta_{34}}%
\frac{G_\calO(u,v)}{(P_{12})^{\frac12(\Delta_{1}+\Delta_{2})}(P_{34})^{\frac12(\Delta_{3}+\Delta_{4})\,%
}}\,,
\eeq
where 
\begin{equation}
u=\frac{P_{12}P_{34}}{P_{13}P_{24}},\quad v=\frac{P_{14}%
P_{23}}{P_{13}P_{24}}.
\end{equation}

For a spinning four point function, partial waves will be classified by the couplings ($\equiv$ three point functions) of the exchanged field to the external fields.  Note that here we are assuming that the exchanged field is a traceless symmetric tensor.  Let us consider these couplings in the differential basis \reef{Dbasis}:
\beq
\left\{ {\scriptsize
\begin{array}{ccc}
\Delta_1   & \Delta_2 & \Delta_0 \\
l_1 & l_2 & l_0 \\
n_{20} & n_{10} & n_{12} \end{array} }
\right\}\,, \qquad
\left\{ {\scriptsize
\begin{array}{ccc}
\Delta_3   & \Delta_4 & \Delta_0 \\
l_3 & l_4 & l_0 \\
n_{40} & n_{30} & n_{34} \end{array} }
\right\}\,,
\label{eq:coupl}
\eeq
where here the exchanged field is denoted by the index 0, and we are considering as before the (12)(34) channel.
In order to instead use the standard three point function structures (\ref{eq:general3pt}) one has to perform a change of basis as explained in section~\ref{sec:recursion}.

The conformal partial wave corresponding to the couplings \reef{eq:coupl} is then given by:
\beq
\boxed{
\vphantom{\sum}\ {\cal D}_{\rm left}{\cal D}_{\rm right}W_{\calO}(P_1,P_2,P_3,P_4)\,,
}
\label{eq:main}
\eeq
where  
\beq
{\cal D}_{\rm left}=
H_{12}^{n_{12}}
{D}_{12}^{n_{10}}
{D}_{21}^{n_{20}}
{D}_{11}^{m_1}
{D}_{22}^{m_2}\,\Sigma^{m_1+n_{20} +n_{12},\,m_2+n_{10} +n_{12}}\,,
\eeq
and ${\cal D}_{\rm right}$ has the same form with $1\to3$ and $2\to 4$. These operators are applied to $W_{\calO}$, the conformal partial wave corresponding to the exchange of the same field $\cal O$ between scalar fields. The symbol $\Sigma^{a,b}$ is present to remind us that dimensions of external scalars in $W_{\calO}$ have to be shifted as in the RHS of \reef{Dbasis}:
\beq
\Delta_1\to\Delta_1+m_1+n_{20} +n_{12}\,,\qquad \Delta_2\to\Delta_2+m_2+n_{10} +n_{12}\,.
\eeq
The shifts for $\Delta_{3,4}$ are given by the same equations with $1\to3$ and $2\to 4$.

\subsubsection{Comments on the Final Result}
The expression~\reef{eq:main} is the main result of this paper. It gives a compact embedding space expression for the conformal blocks. If desired, these could in principle be projected to the physical space.  We would now like to comment on the form of the result and on how it can be used in the future, in particular in the conformal bootstrap program. As we will see, for these applications projection to the physical space is unlikely to be necessary.

To begin with, let us imagine that we have performed the differentiations in Eq.~\reef{eq:main}. In practical applications this will be straightforward to do, particularly due to the fact that interesting cases will involve external spins $\leq 2$. The result will then have the structure expected from a conformal four point function in the embedding space:
 
\beq
 \left( \frac{P_{24}}{P_{14}}\right)^{\frac{\tau_1 - \tau_2}{2}} \left( \frac{P_{14}}{P_{13}}\right)^{\frac{\tau_3 - \tau_4}{2}}
\frac{
  \sum_k  f_k(u,v) \,Q^{(k)} (\{ P_i;Z_i\})}
  {(P_{12})^{\frac{\tau_1 + \tau_2}{2}} (P_{34})^{\frac{\tau_3 + \tau_4}{2}}}\,,
  \label{eq:4pf}
\eeq
where
$Q^{(k)}$ are explicitly transverse polynomials of $Z_i$ which have degree $l_i$ in $P_i$. Similarly to the three point function case, these polynomials are constructed from the basic building blocks $V_{i,jk}$ and $H_{ij}$
introduced in section \ref{sec:3ptEmb}. We refer readers to section 4.3 of \cite{Emb}, where the problem of constructing and enumerating these polynomials is discussed in detail. 

The coefficient functions $f_k(u,v)$ will be determined by this computation; they will be certain linear combinations of derivatives of scalar conformal blocks $G_\calO(u,v)$ with shifted dimensions. In the next section we consider a simple example where we compute these functions explicitly.

In the conformal bootstrap program, one imposes the equality of the expansions in the conformal partial waves in the (12)(34), (13)(24), and (14)(23) channels. A moment's thought shows that this crossing symmetry constraint will be easy to impose using directly the embedding space expressions \reef{eq:4pf}. 

An important role in existing applications of the conformal bootstrap~\cite{us,david,us-global,Vichi,Poland:2011ey} was played by positivity constraints on the coefficients appearing in the conformal partial wave expansion~\reef{eq:CBexp}. In the simplest case of four identical real scalars $\phi_i=\phi$, 
these constraints come from the fact that the coefficients in question are the squares of real OPE coefficients:
\beq
\lambda_{12 \calO}\lambda_{34 \calO}\equiv \lambda_O^2>0\,.
\eeq
For the spinning case there are several (let's say $N_\calO$) possible couplings, so that the diagonal coefficients will be positive, while the cross term coefficients are not sign-definite. The most general requirement in this case is that the whole matrix of $N_\calO \times N_\calO$ coefficients be positive-definite; see an analogous discussion in section 2.5 of \cite{us-global}.

Let us now comment on the `missing' conformal blocks of antisymmetric (and mixed symmetry) fields that we have not yet computed, mentioned at the end of section \ref{sec:basic}. First of all, there are important partial cases when such fields simply do not appear in the OPE.  The OPE of two scalars is one example, but perhaps not the only one.  It is worth checking if there are other special situations (e.g., involving conserved operators) where such fields cannot make an appearance.

Second, in $d=3$ dimensions the antisymmetric fields can be converted into symmetric ones using the $\epsilon$-tensor. In this case we can actually compute all of the conformal partial waves in terms of the scalar ones. The construction requires discussing parity odd three point functions; see section \ref{sec:odd} below.

In all the other cases the antisymmetric and mixed conformal partial waves need to be computed before attempting the bootstrap.
An efficient way to do so may be as follows. For any given symmetry of the exchanged field, one can choose the simplest 
external fields which can couple to it. For example, for an antisymmetric rank 2 tensor $F_{[ab]}$, this is a scalar and a vector, with the OPE
\beq
A_a(x) \times \phi(0)\sim (x^2)^{\frac 12(\Delta_F-\Delta_A-\Delta_\phi+1)} x^b F_{[ab]}\,.
\eeq
Since in this simplest case there is only one possible coupling, there will be only one conformal partial wave, and it can perhaps be found using the Casimir differential equation method of section~\ref{sec:casimir} (compare with footnote~\ref{note:3}). If this is done, then external spin can be increased recursively, just as we did in this paper. 

\subsubsection{Example}

Let us now exemplify the method concretely, by determining the conformal blocks for the four point function of two spin 1 operators at positions 1 and 2 and two scalar operators at positions 3 and 4.
We will present the conformal blocks corresponding to the differential basis (\ref{Dbasis}) of three point functions. If one wishes, one can easily convert to the standard basis (\ref{eq:general3pt}) using the discussion of section \ref{sec:example3pt}.
There are then 5 independent conformal partial waves (for internal spin greater than 1): 
\begin{align}
D_{11}D_{22} &\,W_\calO^{1,1}(P_1,P_2,P_3,P_4)\,, \nn\\
D_{21}D_{11} &\,W_\calO^{2,0}(P_1,P_2,P_3,P_4)\,, \nn\\
D_{12} D_{22} &\,W_\calO^{0,2}(P_1,P_2,P_3,P_4)\,, \label{eq:list5}\\
D_{12}D_{21} &\,W_\calO^{1,1}(P_1,P_2,P_3,P_4)\,, \nn \\
H_{12} &\,W_\calO^{1,1}(P_1,P_2,P_3,P_4)\,, \nn
\end{align}
where the superscript in $W_\calO^{k_1,k_2}$ denotes the integer shifts in the dimensions of the external operators $\Delta_1\to \Delta+k_1$ and $\Delta_2\to \Delta_2+k_2$. 

As mentioned in the previous section, it is possible to represent these partial waves in the form \reef{eq:4pf}. In this case there are five polynomials $Q^{(k)}$ representing tensor structures (see section 4.3.1 of \cite{Emb}). One possible basis is given by:
\beq
Q^{(k)}=\{ V_{1,23}V_{2,13}\,,\ V_{1,24}V_{2,14}\,,\ V_{1,23}V_{2,14}\,,\ V_{1,24}V_{2,13}\,,\ H_{12} \}\,.
\eeq
For illustrative purposes, we give here the coefficient functions for the first partial wave in the list \reef{eq:list5}, specializing to the case $\Delta_1=\Delta_2$, $\Delta_3=\Delta_4$ for simplicity:
\begin{gather}
f^{(1)}=f^{(2)}=
 u\, w\, \del_u \del_{w}\, G_\calO\,,\nn \\
f^{(3)}=(w\, \del_{w})^2 G_\calO\,,\qquad 
f^{(4)}= (u\, \del_u)^2 G_\calO\,,  \\
f^{(5)}= -\frac{1}{2} (u\, \del_u +w\, \del_{w}) G_\calO\,. \nn
\end{gather}
Here $G_\calO$ is the scalar conformal block with external dimensions $(\Delta_1+1,\Delta_2+1,\Delta_3,\Delta_4)$,\footnote{Actually, the scalar conformal blocks only depend on the differences $\Delta_1-\Delta_2$ and $\Delta_3-\Delta_4$ of the conformal weights of the external operators.}
expressed in terms of $u$ and $w \equiv u/v$. Under 
$P_3\leftrightarrow P_4$ we have 
\beq
Q^{(1)}\leftrightarrow Q^{(2)},\  Q^{(3)}\leftrightarrow Q^{(4)},\ Q^{(5)}=\text{inv},\qquad u\leftrightarrow w\,,
\eeq  
which expains the symmetry visible in the given expressions for $f^{(k)}$.  

\subsection{Special Cases}

In the previous section we presented our method of computing spinning conformal partial waves in the most basic, generic case. There exist three special situations when the method needs to be modified or expanded; they are treated in this section.

\subsubsection{Three Dimensions}

The case of three spacetime dimensions is special, because the classification of three point functions as presented in section
\ref{sec:3ptEmb} has to be modified. Namely, not all three point functions of the form \reef{eq:general3pt} are independent. This is because for $d=3$ there exist a degeneracy between the elementary building blocks, expressed by the identity \cite{Emb}\footnote{The problem of classifying $d=3$ three point functions without overcounting was first solved in \cite{Giombi} using a different language.}
\beq
\left(V_1H_{23}+V_2H_{13}+V_3H_{12}+2V_1V_2V_3\right)^2= -2H_{12}H_{13}H_{23} +O(Z_i^2, Z_i\cdot P_i)\,.
\label{evenidentity}
\eeq
This identity is a consequence of the fact that six $(d+2)$-dimensional vectors $Z_i$, $P_i$ must be linearly dependent for $d=3$.

By means of this identity, we can take as truly independent three point functions those for which at least one of three parameters $n_{ij}$ vanishes. The counting formula for these structures is given in \cite{Emb}.

There is little change to the rest of the algorithm of computing partial waves. We have to represent the independent three point functions via the differential operators \reef{eq:diffops} and \reef{eq:diffops2} acting on the scalar three point functions. One can do this by using the recursion relations, or by using the matrix inversion method as described at the end of section \ref{sec:recursion} and illustrated in section \ref{sec:example3pt}. The matrix $a_{IJ}$ expressing the elements of the \emph{full} differential basis in terms of the \emph{reduced} standard basis will now be rectangular. This just means that the elements of the differential basis will also be linearly dependent for $d=3$. Unlike for the standard basis, we were unable to find a general way to give a canonical linearly independent basis of differential structures. In practice, one can just pick a square submatrix of $a_{IJ}$ and invert it. The existence of the recursion relations guarantees that $a_{IJ}$ is of full rank and that an invertible square submatrix must exist.

Once a differential representation is known, one computes the conformal blocks as in section \ref{sec:main}.

\subsubsection{Parity Odd Three Point Functions}
\label{sec:odd}
So far we have discussed the case of parity even three point functions.
This is sufficient in $d>4$ where this is always the case.
However, in $d=4$ one can use the $\epsilon$-tensor to make parity odd conformally invariant three point functions. As explained in \cite{Emb}, any  parity odd structure can be obtained by multiplying parity even structures (\ref{eq:general3pt}) by
\beq
\epsilon(Z_1,Z_2,Z_3,P_1,P_2,P_3)\,.
\eeq
Here and below $\epsilon(\ldots)$ denotes the full contraction of the $\epsilon$-tensor with the shown vectors.

We would like to be able to generate these structures by acting with differential operators (that only depend on $P_1,P_2,Z_1,Z_2$) on the three point function of two scalars at points 1 and 2 and an operator with spin at point 3.
This goal can be achieved with the help of the differential operator
\beq
\epsilon\left(Z_1,Z_2, P_1,P_2,\frac{\partial}{\partial P_1},\frac{\partial}{\partial P_2}\right)\,. \label{eq:eps4d}
\eeq
This operator satisfies the same two consistency conditions as the operators $D_{ij}$. First, it is interior to the surface defined by the constraints \reef{eq:surf}, i.e.\ the result of its action on a function $F(\{P_i;Z_i\})$ is independent of the values of $F$ outside the constraint surface.
Second, it preserves the explicit transversality of $F$: $F(\{P_i;Z_i+\beta_i P_i\})=F(\{P_i;Z_i\})$ for all $\beta_i$.
Therefore, we can use this operator to generate all parity odd three point functions by acting on the parity even three point functions (\ref{Dbasis}).
Note that the order is irrelevant because (\ref{eq:eps4d}) commutes with $H_{12}$ and all $\mathcal{D}_{ij}$.

In $d=3$ the situation is slightly more complicated. In this case, there are three parity odd building blocks
\beq
\epsilon_{ij}\equiv P_{ij}\,  \epsilon(Z_i,Z_j,P_1,P_2,P_3)\qquad(i,j=1,2,3)\,, 
\eeq
which are however not independent \cite{Giombi,Emb}.
We will not attempt to give a complete classification of all possible three point functions in terms of differential operators acting on the basic structure, as in Eq.~(\ref{Dbasis}).
Instead, we will define differential operators that can be used to generate any given parity odd three point function. The following operators are interior and preserve transversality:
\begin{align}
\tilde{D}_1&=\epsilon\left(Z_1,   P_1,\frac{\partial}{\partial P_1},P_2,\frac{\partial}{\partial P_2}\right)
+\epsilon\left(Z_1,   P_1,\frac{\partial}{\partial P_1},Z_2,\frac{\partial}{\partial Z_2}\right)\nonumber\label{eq:eps3d}\,,\\
\tilde{D}_2&=\epsilon\left(Z_2,   P_2,\frac{\partial}{\partial P_2},P_1,\frac{\partial}{\partial P_1}\right)
+\epsilon\left(Z_2,   P_2,\frac{\partial}{\partial P_2},Z_1,\frac{\partial}{\partial Z_1}\right) \,. 
\end{align}
The first (second) operator adds one unit of spin at point $P_1$ ($P_2$) without changing the conformal dimensions.
One can also define interior and transversality preserving operators that add spin at both points at once:
\begin{align}
\epsilon\left(Z_1,  Z_2, P_1, P_2,\frac{\partial}{\partial P_1}\right)\, ,\ \ \ \ \ \ \ \ 
\epsilon\left(Z_1,  Z_2, P_1, P_2,\frac{\partial}{\partial P_2}\right)\, .
\end{align}
These operators may then be used to generate any given parity odd structure in three dimensions.

\subsubsection*{Example:}

Let us exemplify the method for parity odd structures in $d=3$ in the particular case discussed in subsection \ref{sec:example3pt}.
In this case, there are 4 independent parity odd tensor structures that can be chosen as\footnote{We use a tilde to distinguish these from the parity even structures of subsection \ref{sec:example3pt}.}
\begin{align}
\tilde{[1]}\equiv\epsilon_{23} \left[ {\scriptsize
\begin{array}{ccc}
\Delta_1+1 & \Delta_1+2 & \Delta+2 \\
1 & 0 & l-1 \\
0 & 0 & 0 \end{array} }
\right],\quad
\tilde{[2]}\equiv\epsilon_{23} \left[ {\scriptsize
\begin{array}{ccc}
\Delta_1+1 & \Delta_1+2 & \Delta+2 \\
1 & 0 & l-1 \\
0 & 1 & 0 \end{array} }
\right],\nn\\
\tilde{[3]}\equiv\epsilon_{13} \left[ {\scriptsize
\begin{array}{ccc}
\Delta_1+2 & \Delta_1+1 & \Delta+2 \\
0 & 1 & l-1 \\
0 & 0 & 0 \end{array} }
\right],\quad
\tilde{[4]}\equiv\epsilon_{13} \left[ {\scriptsize
\begin{array}{ccc}
\Delta_1+2 & \Delta_1+1 & \Delta+2 \\
0 & 1 & l-1 \\
1 & 0 & 0 \end{array} }
\right].
\end{align}
For $l\ge 2$, the third parity odd structure $\epsilon_{12}$ can be always eliminated in favor of $\epsilon_{13}$ and $\epsilon_{23}$ by means of the $d=3$ identity \cite{Emb}
\beq
V_3^2 \epsilon_{12}=(H_{13}+V_1V_3)\epsilon_{23}-
(H_{23}+V_2V_3)\epsilon_{13}\,.
\label{eq:id1}
\eeq

Alternatively, we can use the differential operators (\ref{eq:eps3d})
to write another basis
\begin{align}
\tilde{\{1\}}\equiv\tilde{D}_1 D_{21} \left[ {\scriptsize
\begin{array}{ccc}
\Delta_1+1 & \Delta_1 & \Delta \\
0 & 0 & l \\
0 & 0 & 0 \end{array} }
\right],\quad
\tilde{\{2\}}\equiv\tilde{D}_1 D_{22} \left[ {\scriptsize
\begin{array}{ccc}
\Delta_1 & \Delta_1+1 & \Delta \\
0 & 0 & l \\
0 & 0 & 0 \end{array} }
\right], \nn\\
\tilde{\{3\}}\equiv\tilde{D}_2 D_{12} \left[ {\scriptsize
\begin{array}{ccc}
\Delta_1 & \Delta_1+1 & \Delta \\
0 & 0 & l \\
0 & 0 & 0 \end{array} }
\right],\quad 
\tilde{\{4\}}\equiv\tilde{D}_2 D_{11} \left[ {\scriptsize
\begin{array}{ccc}
\Delta_1 +1& \Delta_1 & \Delta \\
0 & 0 & l \\
0 & 0 & 0 \end{array} }
\right].
\label{eq:Itilde}
\end{align}
It is possible to find an explicit relation between the two bases:
\beq
\tilde{\{I\}}=\sum_{J=1}^4 \,b_{IJ} \,\tilde{[J]}\,.
\label{eq:a}
\eeq
To find this relation, one implements the action of differential operators in \reef{eq:Itilde} and reduces the result to a linear combination of the basis $\tilde{[J]}$ 
by identities which replace all occurring $\epsilon$-tensor contractions by $\epsilon_{13}$ and $\epsilon_{23}$. Concretely, in addition to \reef{eq:id1} mentioned above, one uses the identity\footnote{This identity is obtained by expanding the first line in the $6\times6$ determinant
\beq
\left|
\begin{array}{cccccc}
A_1& A_2 & A_3 & A_4 & A_5 & A_6 \\
Z_1 & Z_2 & Z_3 & P_1 & P_2 & P_3 
\end{array} 
\right|=0\,,
\eeq
made to vanish by picking the numbers $A_i$ to respect the linear dependence relation which necessarily exists for the six 5-vectors forming the last five lines. See Eq.\ (4.50) of \cite {Emb} where the same idea was applied in the physical space.}
\begin{align}
   2 (P_1\cdot P_2)&(P_1\cdot P_3)\epsilon
   \left(P_2,P_3,Z_1,Z_2,Z_3\right)\nonumber\\
&=\left((P_1\cdot P_3)(P_2\cdot
   Z_3)-(P_2\cdot P_3)(P_1\cdot
   Z_3)\right) \epsilon
   \left(P_1,P_2,P_3,Z_1,Z_2\right)\nonumber\\
  &+
   \left((P_2\cdot P_3)(P_1\cdot
   Z_2)-(P_1\cdot P_2)(P_3\cdot
   Z_2)\right) \epsilon
   \left(P_1,P_2,P_3,Z_1,Z_3\right)\nonumber\\&+
   \left((P_1\cdot P_3)(P_2\cdot
   Z_1)+(P_1\cdot P_2)(P_3\cdot
   Z_1)\right) \epsilon
   \left(P_1,P_2,P_3,Z_2,Z_3\right)\,,
\end{align}
as well as two similar identities (related by permutations), allowing us to eliminate the structures $\epsilon \left(P_i,P_j,Z_1,Z_2,Z_3\right)$.

The matrix $b_{IJ}$ is then given by
\beq
(\Delta-1)(\Delta_1-1) \left(
\begin{array}{cccc}
 -1-\frac{l}{\Delta-1} &
   -1+\frac{l}{\Delta_1-1}  &
   (1-\frac{l}{\Delta-1}) (1+ \frac{l(\Delta
   -1)}{\Delta_1-1}) & 1 -\frac{l^2}{\Delta_1-1}
   \\
   -1+\frac{l}{\Delta-1}-\frac{2l}{\Delta_1-1}
   & -1- \frac{l}{\Delta_1-1} &
   (1+\frac{l}{\Delta-1}) (1+\frac{l (\Delta
   -1)}{\Delta_1-1}) &  1+\frac{l^2}{\Delta _1-1}
   \\
 (-1+\frac{l}{\Delta-1}) (1+\frac{l (\Delta
   -1)}{\Delta _1-1}) & -1+\frac{l^2}{\Delta_1-1}
   & 1+\frac{l}{\Delta-1}  & 1-\frac{l}{\Delta _1-1} \\
 (-1-\frac{l}{\Delta-1}) (1+\frac{l (\Delta
   -1)}{\Delta_1-1}) & -1-\frac{l^2}{\Delta_1-1}
   & 1-\frac{l}{\Delta-1} + \frac{2l}{\Delta_1-1} &
    1+\frac{l}{\Delta_1-1}
\end{array}
\right)\,.
\eeq
Note that the determinant 
\beq
\det b_{IJ}= 4 l^4 (l^2-1) (\Delta -2) (\Delta -1)^4 \Delta 
\eeq
is positive for $l\ge2$, as $\Delta\ge 3$ is required in this case by the $d=3$ unitarity bound. This shows that the differential structures \reef{eq:Itilde} indeed provide a basis for the parity odd three point functions.

\subsubsection{Conserved Tensors}

Now we will discuss the situation when some of the operators appearing in the four point function are {\it conserved} spin $l$ operators, having dimensions $\Delta = l + d - 2$ that saturate the $d$-dimensional unitarity bound.  Conservation requires that some of the three point function coefficients appearing in the sum Eq.~(\ref{eq:embsum}) are related to each other, and these constraints then appear as relations between the coefficients appearing in the conformal block decomposition Eq.~(\ref{eq:4pf}).  

In~\cite{Emb} we showed that these constraints could be efficiently analyzed in the embedding formalism.  One does this by requiring that the action of the operator $\partial_P \cdot D_Z$ vanishes when applied to Eq.~(\ref{eq:embsum}), for a conserved operator at the point $\{P,Z\}$, where
\beq
\partial_P \cdot D_Z \equiv \frac{\partial}{\partial P_M} \left[\left(\frac{d}{2} - 1 + Z \cdot \frac{\partial}{\partial Z} \right) \frac{\partial}{\partial Z^M} - \frac12 Z_M \frac{\partial^2}{\partial Z \cdot \partial Z} \right]\,.
\eeq
These constraints were studied in the basis of $V_i$'s and $H_{ij}$'s in~\cite{Emb}; let us now take a moment to see what these constraints look like in the differential operator basis in a simple example.

\subsubsection*{Example:}

In the case of two conserved spin $1$ currents at points $1$ and $2$, and a spin $l$ operator of dimension $\Delta$ at point $3$, we can use the basis of 5 parity even structures $\{I\}$ introduced in subsection~\ref{sec:example3pt}. Requiring that $\partial_{P_i} \cdot D_{Z_i}$ for $i=1,2$ vanishes on an arbitrary parity even three point function structure $\sum_{I = 1}^5 \alpha_I \{I\}$ then leads to the 3 constraints
\begin{align}
\alpha_1 &= -\frac{(\Delta-l-d)(\Delta+l-2) \alpha_2 - 2(d-2) \alpha_4 + 4 \alpha_5}{C_{\Delta,l}}\,, \\
\alpha_2 &= \alpha_3 = -\frac{(\Delta+l)(\Delta-l-d+2)}{C_{\Delta,l}} \alpha_4\,,
\end{align}
which reduces the number of independent structures down to 2. We do not know of a natural reason why the Casimir eigenvalue \reef{eq:caseig} arises in these expressions.

Similarly, we can impose conservation on the general parity odd three point function  $\sum_{I = 1}^4 \beta_I \tilde{\{I\}}$ discussed in the previous section for $d=3$. The structures $\tilde{\{1\}}$ and $\tilde{\{2\}}$ are automatically conserved at $P_1$, and the structures $\tilde{\{3\}}$ and $\tilde{\{4\}}$ are automatically conserved at $P_2$. Conservation at both points $P_1$ and $P_2$ requires
\begin{align}
\frac{\beta _2}{\beta _1}= \frac{\beta _4}{\beta _3}=-\frac{(\Delta -l-2)
   (\Delta +l-1)}{
   C_{\Delta,l}+2}\, ,
   \end{align}
reducing the number of independent parity odd structures from 4 to 2.  

Constraints on the three point function coefficients of other conserved operators in the differential operator basis can be worked out similarly.

\subsection{Conformal Blocks for Conserved Currents in CFT$_3$  } \label{sec:3dcurrentsCB}
All irreducible representations of $SO(3)$ are totally symmetric traceless tensors. Therefore, our method is sufficient to determine all of the conformal blocks in $d=3$ dimensions.
In this section, we present explicit expressions for the important case of conserved currents.

The $d=3$ conformal partial waves for a four point function of conserved currents are given by the main formula (\ref{eq:main}), where the operators $\mathcal{D}_{\rm left}$ and 
$\mathcal{D}_{\rm right}$ can take one of 4 possible forms.
This gives a total of $4\times 4=16$ conformal blocks for each primary operator appearing in the OPE of the currents.
From these 4 possible three point functions between two conserved currents and a generic  operator with spin $l\ge2$, two are parity even and two are parity odd.
Using the results of the previous section, the parity even possibilities for 
$\mathcal{D}_{\rm left}$ can be written as
\begin{align}
\mathcal{D}^{(1)}_{\rm left}=&\left( 
2+\frac{(\Delta -l-1) (\Delta -l-3) (\Delta
   +l-2) (\Delta +l)}{C_{\Delta,l}}\right)
   D_{11}D_{22}\Sigma^{1,1}\nn\\
   &- 
   (\Delta -l-1) (\Delta +l) 
   \left(D_{21}D_{11}\Sigma^{2,0}
   +D_{12}D_{22}\Sigma^{0,2}\right)+
  C_{\Delta,l}
  D_{12}D_{21}\Sigma^{1,1}\,,
   \\
\mathcal{D}^{(2)}_{\rm left}=&-4D_{11}D_{22}\Sigma^{1,1}+C_{\Delta,l}
H_{12}\Sigma^{1,1}\,,
\end{align}
while the parity odd forms are
\begin{align}
\mathcal{D}^{(3)}_{\rm left}&= \left(C_{\Delta,l}+2\right)\tilde{D}_1D_{21} \Sigma^{1,0}-
   (\Delta -l-2)
   (\Delta +l-1)\tilde{D}_1D_{22} \Sigma^{0,1}\, ,\\
 \mathcal{D}^{(4)}_{\rm left}&= \left(C_{\Delta,l}+2\right)\tilde{D}_2D_{12} \Sigma^{0,1}-
   (\Delta -l-2)
   (\Delta +l-1)\tilde{D}_2D_{11} \Sigma^{1,0}\,.
\end{align}
As in Eq.~\reef{eq:main}, $\Sigma^{a,b}$ in these expressions denotes the shifts in the dimensions appearing in the scalar conformal block that must be performed before acting with the differential operators.

\section{Conclusions} \label{sec:concl}
In this paper we presented a method to efficiently derive the conformal blocks corresponding to the exchange of traceless symmetric tensors appearing in four point functions of operators with spin.  In particular, we found that all such conformal blocks may be expressed in terms of simple differential operators acting on the basic scalar conformal blocks, given in Eq.~(\ref{eq:main}) for parity even structures.  To obtain this result, we made extensive use of the index-free embedding space formalism developed in~\cite{Emb}.  

Our results may be especially useful when applied to four point functions containing conserved currents or the stress tensor.  For example, it would be extremely interesting to pursue the conformal bootstrap program for four point functions of these important operators, perhaps following the lines of~\cite{us,david,us-global,Vichi,Poland:2011ey}.  It is plausible that studying crossing symmetry of stress tensor four point functions could lead to interesting constraints on central charges in CFTs, perhaps not unrelated to the bounds of~\cite{Hofman:2008ar} or~\cite{Hellerman:2009bu}.  In order to pursue these ideas the results of this paper will be an important ingredient.  

They may also be useful in the context of the AdS/CFT correspondence, where stress tensor four point functions are related to graviton scattering amplitudes.  This connection is particularly transparent when CFT correlators are written in the Mellin representation~\cite{MackMellin,joaoMellin,Fitzpatrick:2011ia,Paulos:2011ie}, and we believe that the formalism developed in this paper will be useful for further exploring this connection.  A possible goal of this program might be to utilize some of the powerful recursion relations known for scattering amplitudes (e.g.,~\cite{Britto:2005fq}) in order to holographically compute CFT correlators of the stress tensor.  Some steps towards this goal were taken in~\cite{Suvrat,Fitzpatrick:2011ia,Paulos:2011ie}; we hope that the present formalism will allow for additional progress to be made.

In this paper, we succeeded at giving expressions for conformal blocks corresponding to the exchange of traceless symmetric tensors, since these are the only operators that can appear in scalar OPEs.  Other possible Lorentz representations, such as antisymmetric tensors or operators with mixed symmetry, will require an alternative approach such as solving the Casimir differential equation.  However, it is likely that once the simplest (i.e., lowest external spin) versions of these blocks could be found, the blocks corresponding to higher external spins could again be computed by applying simple differential operators as in the present paper.  We leave further explorations of this approach to future work.

Finally, we would like to emphasize that our method is sufficient to give all conformal blocks in three dimensional CFTs. This point was exemplified in section \ref{sec:3dcurrentsCB} for the four point function of conserved currents. This result can be easily extended to the case of the stress-energy tensor, providing all the basic ingredients to apply the bootstrap program to CFTs dual to pure four dimensional quantum gravity on AdS$_4$.


\newpage

\begin{center} 
{\bf Acknowledgements} 
\end{center}
We thank Diego Hofman, Hugh Osborn, David Simmons-Duffin and Pedro Vieira for helpful comments and conversations.  This work was funded in part by the research 
grants PTDC/FIS/099293/2008 and CERN/FP/109306/2009. \emph{Centro de F\'{i}sica do Porto} is partially funded by FCT.
The work of S.R. was supported in part by the European Program ``Unification in the LHC Era", contract PITN-GA-2009-237920 (UNILHC).  S.R. is additionally grateful to the Perimeter Institute and to CERN for hospitality.  The work of D.P. was supported in part by the Harvard Center for the Fundamental Laws of Nature and by NSF grant PHY-0556111.  
Research at the Perimeter Institute is supported in part by the Government of Canada through NSERC and by the Province of Ontario through the Ministry of Research \& Innovation.

\end{document}